\documentclass[final,1p,times]{elsarticle} 

\usepackage{graphicx}
\usepackage{amssymb} 
\usepackage{amsthm} 
\usepackage{atlasphysics}

\newcommand{\RFour}{\mbox{$R = 0.4$}}
\newcommand{\RThree}{\mbox{$R = 0.3$}}
\newcommand{\RTwo}{\mbox{$R= 0.2$}}
\newcommand{\Rcp}{\mbox{$R_{\rm CP}$}}

\newcommand{\antikt}{\mbox{anti-\kt}}
\newcommand{\kt}{\mbox{$k_\mathrm{T}$}}
\newcommand{\pt}{\mbox{$p_\mathrm{T}$}}
\newcommand{\Et}{\mbox{$E_\mathrm{T}$}}
\newcommand{\Ettruth}{\mbox{$E_\mathrm{T}^{\rm truth}$}}
\newcommand{\Etreco}{\mbox{$E_\mathrm{T}^{\rm reco}$}}
\newcommand{\sqrtsnn}{\mbox{$\sqrt{s_{\mathrm{NN}}}$}}
\newcommand{\Ncoll}{\mbox{$N_{\mathrm{coll}}$}}

\newcommand{\Npart}{\mbox{$N_{\rm part}$}}
\newcommand{\Nch}{\mbox{$N_{\rm ch}$}}
\newcommand{\Njet}{\mbox{$N_{\rm jet}$}}
\newcommand{\dphi}{\mbox{$\Delta\phi$}}
\newcommand{\fd}{\mathrm{d}}
\newcommand{\Rcpcent}{\mbox{$R_{\rm CP}^{\rm cent}$}}
\newcommand{\Rcollcent}{\mbox{$R_{\rm coll}^{\rm cent}$}}
\journal{Nuclear Physics A}

\begin{document}

\begin{frontmatter} 

\title{Jets in heavy ion collisions with ATLAS}

\author{Martin Spousta (for the ATLAS\fnref{col1} Collaboration)}
\fntext[col1] {A list of members of the ATLAS Collaboration and acknowledgements can be found at the end of this issue.}
\address{Columbia University in the city of New York, Charles University in Prague}


\begin{abstract} 

The energy loss of high-\pt\ partons 
provides insight into the transport properties of the medium
created in relativistic heavy ion collisions. Evidence for this energy
loss was first experimentally established through observation of
high-\pt\ hadron suppression at RHIC.
  More recently, measurements of fully reconstructed jets have been 
performed at the LHC. In this summary the latest experimental 
results from the ATLAS collaboration on jet suppression are 
presented. In particular the jet suppression in inclusive jet 
yields, path length dependence of the jet suppression, photon-jet 
and $Z^{0}$-jet correlations, heavy flavor suppression, and jet 
fragmentation are discussed.
  These results establish qualitative features of the jet 
quenching mechanism as experimental fact and provide constraints on 
models of jet energy loss.

\end{abstract} 

\end{frontmatter} 



\section{Introduction}
\label{sec:intro}

Collisions of lead ions at Large Hadron Collider (LHC) are 
expected to create a strongly interacting matter in which quarks 
and gluons are locally deconfined. Jets produced in hard scattering 
are expected to be modified due to the energy loss of outgoing 
parton traversing the hot QCD medium. This effect is commonly 
referred to as a jet quenching. Jets therefore represent an 
important tool for studying both the properties of the deconfined 
matter and the parton energy loss \cite{jqt}. In this paper we will 
provide a status report on jet measurements performed using the 
2010 and 2011 Pb+Pb data at $\sqrtsnn = 2.76$~\TeV collected by the 
ATLAS detector \cite{Atlas:2008}. 

The first evidence of the jet quenching at LHC has been observed in 
the measurement of the di-jet asymmetry \cite{Aad:2010bu}. In 
central heavy ion collisions an excess of events with large di-jet 
asymmetry has been observed when compared to p+p or Pb+Pb Monte 
Carlo (MC) reference. This asymmetry has been accompanied by a 
balance in azimuth, that is jets in the di-jet system remain 
``back-to-back'' despite to a sizable modification of their energy. 
This measurement is an observation measurement and by itself it is 
insufficient to provide detailed information about the parton 
energy loss. To gain more information we need to ask more detailed 
questions: We need to know how does the modification of the 
inclusive jet spectra depend on the jet \pt, jet size, or direction 
of the jet with respect to the reaction plane. We also need to know 
whether the suppression depends on the flavor of the initial 
parton. Last but not least we should learn how is the internal 
structure of jets modified. These questions will be addressed in 
the sections~\ref{sec:inclusive}-\ref{sec:frag}. Section 
\ref{sec:event} summarizes the event selection and centrality 
definition, section \ref{sec:perform} discusses the jet 
reconstruction performance.

\section{Event selection and centrality definition}
\label{sec:event}

Analyses presented in this summary use 2010 or 2011 LHC heavy 
ion runs, colliding nuclei at $\sqrtsnn = 2.76$~\TeV. The 2010 run 
collected approximately $7~\mu\mathrm{b}^{-1}$ whereas 2011 run collected 
$0.14$~nb$^{-1}$.
  Minimum bias Pb+Pb collisions were identified using measurements 
from the zero-degree calorimeters (ZDCs) and the minimum-bias 
trigger scintillator (MBTS) counters\footnote{The ZDCs are located 
symmetrically at $z = \pm 140$~m from the interaction point and 
cover $|\eta| > 8.3$. In Pb+Pb collisions the ZDCs primarily 
measure ``spectator'' neutrons, that is neutrons from the incident 
nuclei that do not interact hadronically. The MBTS measures charged 
particles over $2.1 < |\eta| < 3.9$.}.
  In the offline analysis, events were required to have a 
primary vertex reconstructed from charged particle tracks with $\pt
> 500$~\MeV. The 2010 data-taking used only the minimum bias 
trigger whereas the in the 2011 run, events were selected for 
recording using a combination of the minimum bias trigger and 
high-level trigger (HLT). The HLT triggered on jets, high-\pt\ 
photons, electrons, and muons.

  The centrality of each heavy ion collision is determined using 
the sum of the transverse energy in all cells in the forward 
calorimeter ($3.1 < |\eta| < 4.9$), at the electromagnetic scale. 
The average number of nucleon-nucleon collisions, \Ncoll , and the 
average number of participating nucleons, \Npart , were calculated 
using the standard MC Glauber model \cite{glauber}.


\section{Jet reconstruction performance}
\label{sec:perform}

Jets discussed in this summary has been reconstructed at the 
calorimeter level using the \antikt\ algorithm 
\cite{Cacciari:2008gp}. The elliptic flow modulated underlying 
event contribution to the reconstructed jets has been subtracted on 
the event-by-event basis. All the discussed measurements have been 
corrected to the particle level by unfolding procedures such that 
they can be directly compared to theoretical models.

\begin{figure}
\centerline{
\includegraphics[width=0.4\textwidth]{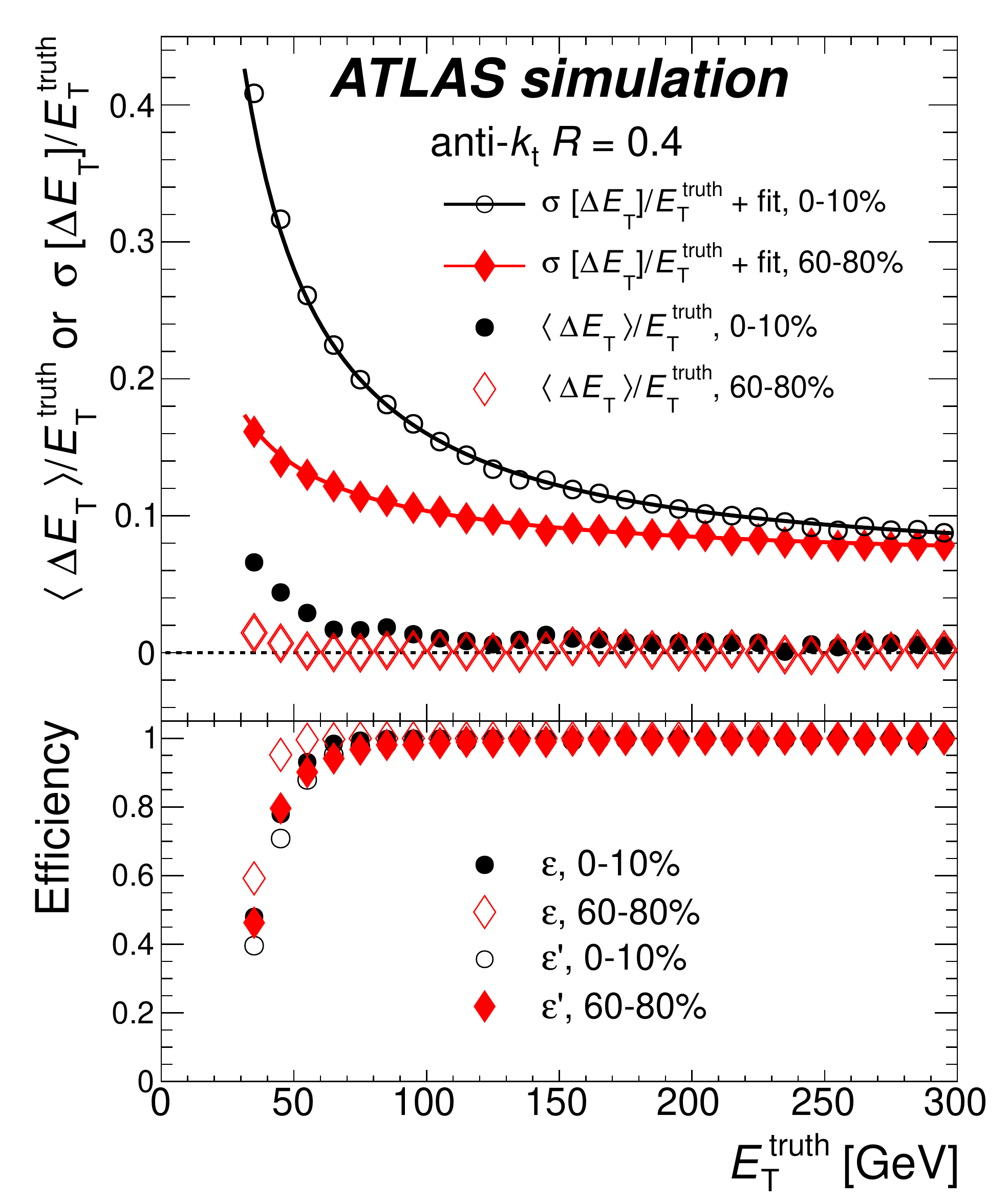} 
\includegraphics[width=0.4\textwidth]{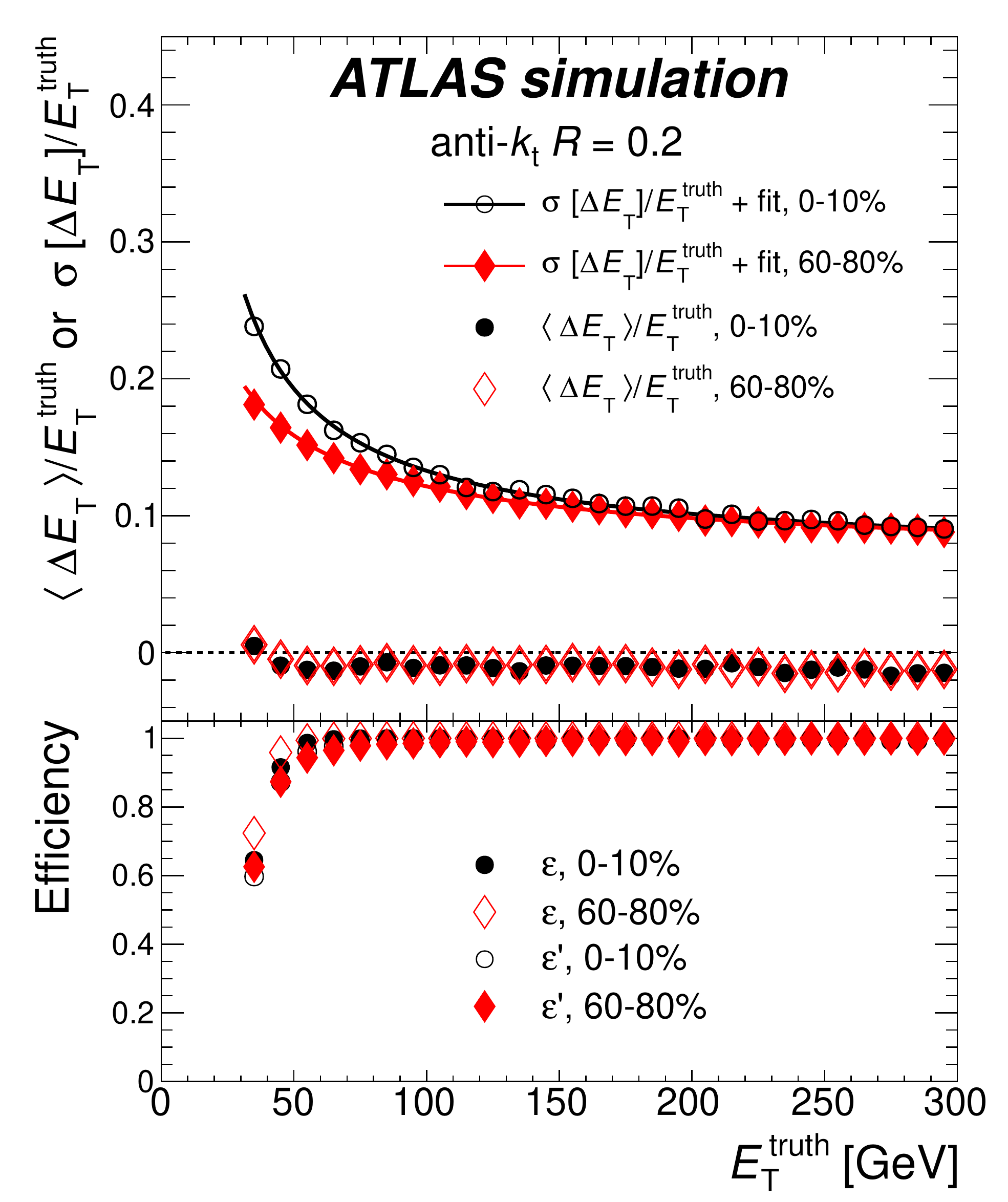}
}
\caption{
  Results of MC evaluation of jet reconstruction performance in 
0-10\% and 60-80\% collisions as a function of truth jet \Et\ for $R 
= 0.2$ (left) and $R = 0.4$ (right) jets \cite{jets}.
  {\it Top}: 
  jet energy resolution $\sigma[\Delta(\Et)]/\Ettruth$ and jet energy 
scale closure $\langle \Delta\Et / \Ettruth \rangle$. Solid curve 
show the parametrization of the JER as described in the text.
  {\it Bottom}: 
  Efficiencies, $\varepsilon$ and $\varepsilon'$, for reconstructing jets 
before and after application of UE-jet removal, respectively.
  }
\label{fig:perform}
\end{figure}

To evaluate the combined performance of the ATLAS detector and the 
jet reconstruction procedures the MC PYTHIA \cite{PYTHIA} di-jet 
events were embedded into HIJING \cite{HIJING} minimum bias events. The 
reference PYTHIA ``truth'' jets were reconstructed using the PYTHIA 
final-state particles by the \antikt\ algorithm with different 
radius parameters of $R=0.2, 0.3, 0.4$, and 0.5. Separately, the 
presence and approximate kinematics of HIJING-generated jets were 
obtained by running $R = 0.4$ \antikt\ reconstruction on final-state 
HIJING particles having $\pt > 4$~\GeV.  To prevent the overlap of 
PYTHIA and HIJING jets from distorting the jet performance evaluated 
relative to PYTHIA truth jets, all PYTHIA truth jets within $\Delta R 
= \sqrt{\Delta\eta^2 + \Delta\phi^2 } = 0.8$ of a $\pt > 10$~\GeV\ 
HIJING jet were excluded from the analysis. 
  To eliminate jets originating from the underlying event 
fluctuations (``UE-jets'') the calorimeter jets were required to 
match a jet reconstructed from inner detector tracks or
  an electromagnetic cluster\footnote{
  Inner detector track jets were reconstructed by the \antikt\ 
$R=0.4$ algorithm from tracks with $\pt > 4$~\GeV\ and were required to 
have $\pt > 7$~\GeV. Electromagnetic clusters were reconstructed from 
cells in the electromagnetic calorimeter and were required to have 
$\pt > 7$~\GeV.
  }.

PYTHIA truth jets passing the HIJING-jet exclusion were matched to 
the closest reconstructed jet of the same $R$ value within $\Delta R 
= 0.2$. The resulting matched jets were used to evaluate the jet 
energy resolution (JER) and the jet energy scale (JES). The jet 
reconstruction efficiency is defined as the fraction of truth jets 
for which a matching reconstructed jet is found. The efficiency was 
evaluated both prior to ($\varepsilon$) and following ($\varepsilon'$) 
UE-jet rejection described above. Figure~\ref{fig:perform} shows a 
summary of the jet reconstruction performance for \antikt\ $R=0.2$ 
and $R=0.4$ jets.

The JER was characterized by $\sigma[\Delta(\Et)]/\Ettruth$, where 
$\sigma[\Delta(\Et)]$ is the standard deviation of the $\Delta \Et = 
\Etreco - \Ettruth$ distribution. 
  The JES closure was evaluated from the mean fractional energy 
shift, $\langle \Delta\Et / \Ettruth \rangle$. The JER was found to 
be well described by a quadrature sum of three terms,

\begin{equation}
\sigma[\Delta(\Et)]/\Ettruth = a/ \kern -0.3em \sqrt{\Ettruth} ~~ \oplus ~~ b/\Ettruth ~~ \oplus ~~ c,
\end{equation}
  where $a$ and $c$ represent the usual sampling and constant 
contributions to calorimeter resolution. The term containing $b$ 
describes the contribution of the underlying event fluctuations to 
the JER. 
  This contribution can be independently evaluated by measuring the 
transverse energy fluctuations in the minimum bias events. This has 
been done in the fluctuation analysis reported in 
Ref.~\cite{fluctuations}. A good consistency between the results of 
the fluctuation analysis and the parametrization of the JER has been 
achieved. Simultaneously, a good agreement between the fluctuations 
in the data and HIJING has been observed.


The jet reconstruction efficiency decreases with decreasing jet \Et\ 
for $\Et \lesssim 50$~\GeV. The decrease starts at larger \Et\ and 
decreases more rapidly for larger jet radii and in more central 
collisions. An independent check of the basic performance has been 
done by using PYTHIA jets embedded directly to the minimum bias data. 
  The reconstruction efficiency was found to differ between central 
and peripheral collisions by less then 5\% on the rise of the 
efficiency curve for $R=0.4$ jets and the agreement in the JES was 
found to be better then a 1\% between central and peripheral 
collisions.
  More details about the jet reconstruction procedures as well as the 
jet performance can be found in Ref.~\cite{jets}.

\begin{figure}
\centerline{
\includegraphics[width=0.4\textwidth]{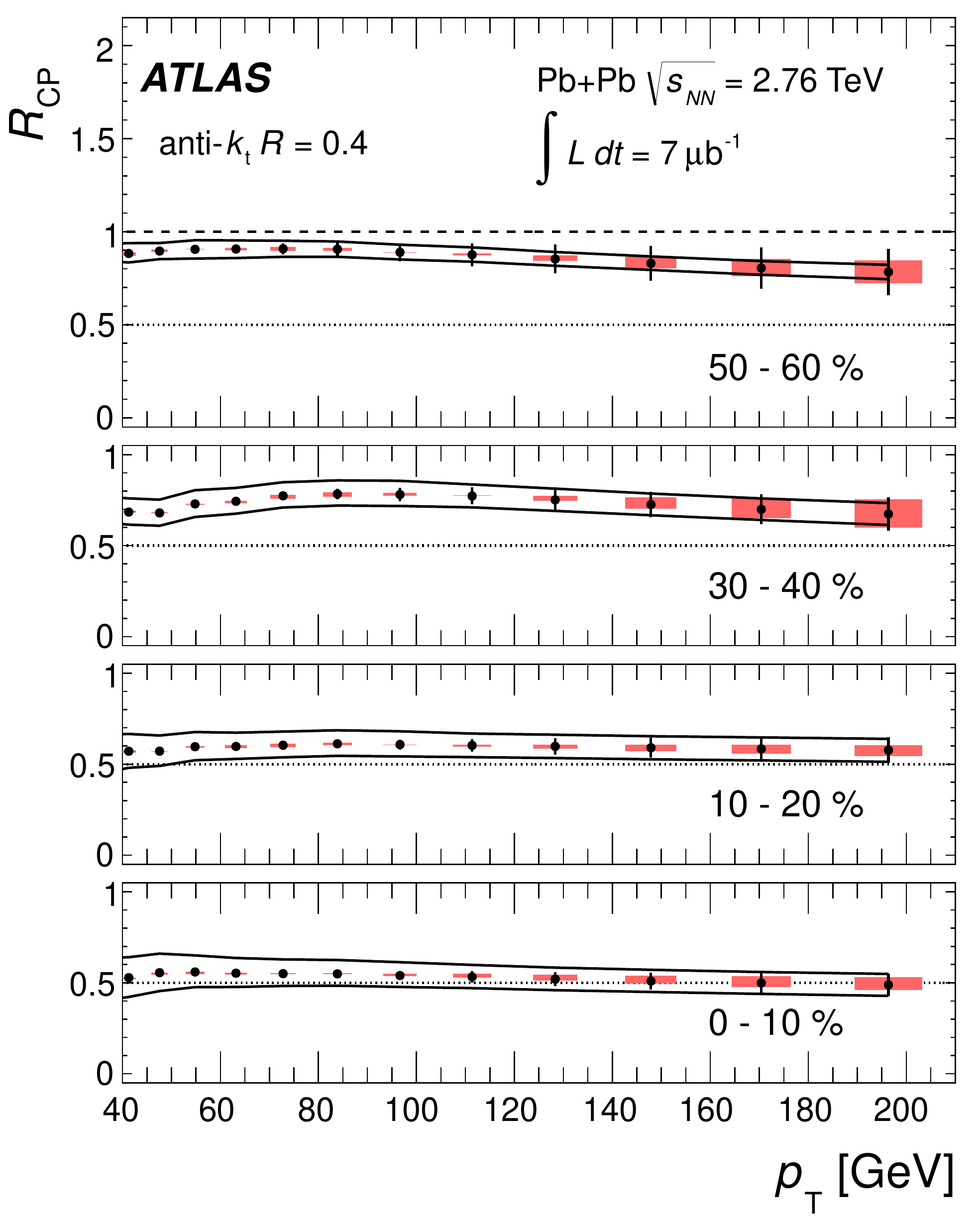} 
\includegraphics[width=0.5\textwidth]{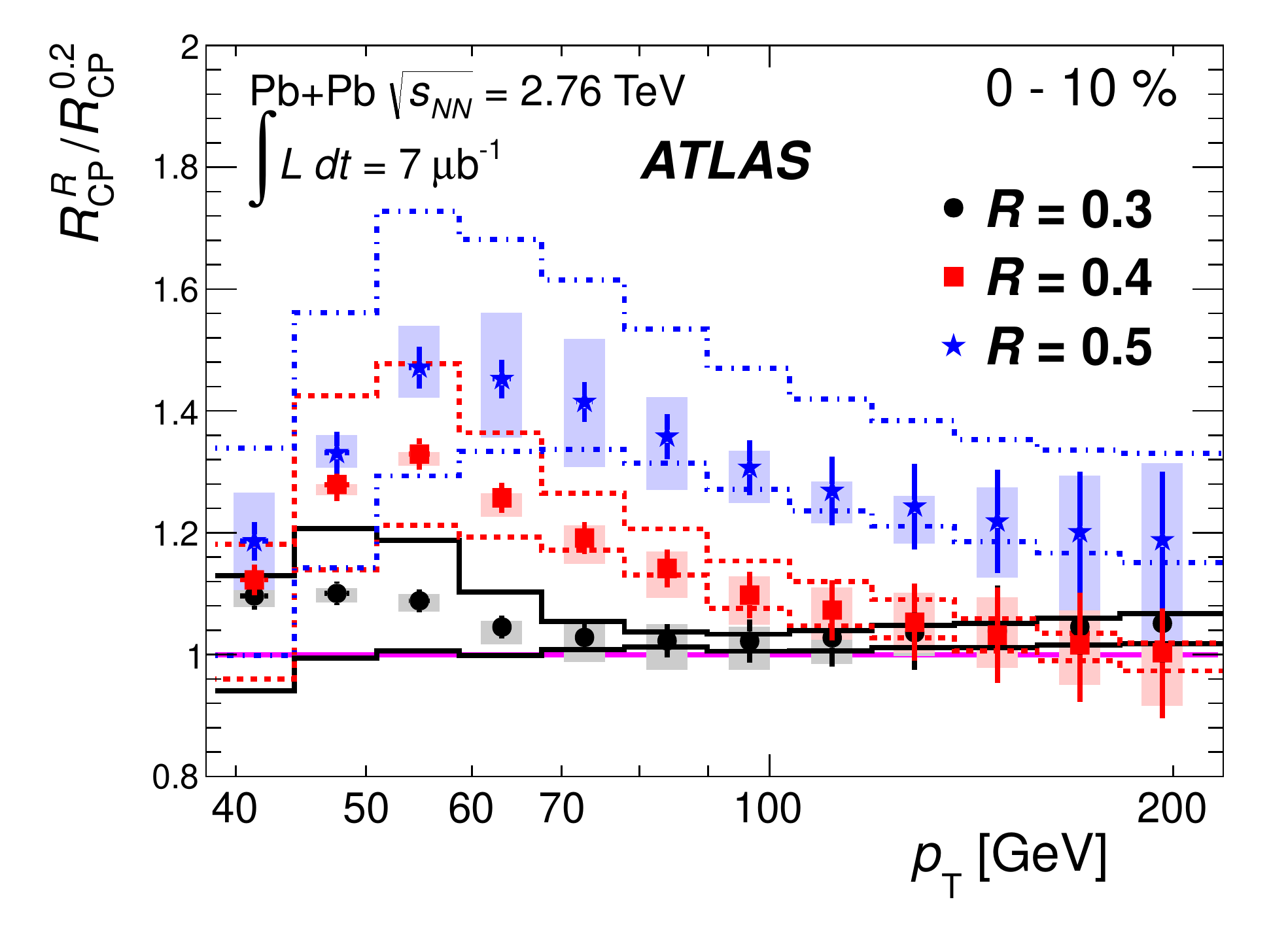}
}
\caption{
  {\it Left:} Jet \Rcp\ values as a function of jet \pt\ for $R=0.4$ 
\antikt\ jets in two bins of collisions centrality. {\it Right:} 
Ratios of jet \Rcp\ values between $R = 0.3, 0.4$, and 0.5 jets and $R 
= 0.2$ jets as a function of \pt\ in the 0-10\% centrality bin. In 
both panels, the error bars show statistical uncertainties, shaded 
boxes indicate partially correlated systematic errors. The lines 
indicate systematic errors that are fully correlated between different 
\pt\ bins \cite{jets}.
  }
\label{fig:jetrcp}
\end{figure}

\section{Inclusive jet suppression}
\label{sec:inclusive}

The first step beyond the original asymmetry measurement is to measure 
the suppression in the inclusive jet yields. To quantify the 
suppression we introduce the jet \Rcp , defined as a fraction of 
per-event normalized jet yield, $\hat{N}$, measured in a given centrality bin to 
the per-event normalized jet yield measured in the 60-80\% most 
peripheral collisions, that is 

\begin{equation}
\Rcpcent = 
(1/\Rcollcent) (\hat{N}^\mathrm{cent}/\hat{N}^{60-80}),
\end{equation}
   where \Rcollcent\ is the central-to-peripheral ratio of 
number of binary collisions. The jet \Rcp\ has been measured for six 
centrality bins, $\mathrm{cent} \in$ \{0-10\%, 10-20\%, 20-30\%, 
30-40\%, 40-50\%, 50-60\%\}. The resulting jet \Rcp\ evaluated as a 
function of jet \pt\ for 0-10\% central and 50-60\% peripheral 
collisions is presented in the left panel of Figure~\ref{fig:jetrcp}. 
One can see that the jet \Rcp\ in the most central collision does not 
exhibit any strong dependence on the jet \pt . A suppression 
by a factor of two of inclusive jet yields in 0-10\% central with 
respect to 60-80\% peripheral collisions can be observed.

The dependence of the jet suppression on the jet size is summarized in 
the right panel of Figure~\ref{fig:jetrcp} in terms of the ratio of 
\Rcp\ values between $R = 0.3, 0.4,$ and 0.5 jets and $R = 0.2$ 
jets, $\Rcp^{R} / \Rcp^{0.2}$, as a function of \pt\ for the 0-10\%
centrality bin. This result indicates a significant dependence of 
\Rcp\ on the jet radius. For $\pt < 100$~\GeV\ the $\Rcp^{R} / \Rcp^{0.2}$ 
values for both $R = 0.4$ and $R = 0.5$ differ from one beyond the 
statistical and systematic uncertainties. This deviation persists well 
beyond 100 \GeV\ and is also present in 10-20\% centrality bin. However, the 
direct comparisons of \Rcp\ between different jet radii at low \pt\ 
should be treated with care as the same jets measured using smaller 
radii will tend to appear in lower \pt\ bins than when measured with a 
larger radius. Further details on the inclusive jet suppression can be found 
in Ref.~\cite{jets}.

\begin{figure}
\centerline{
\includegraphics[width=0.85\textwidth]{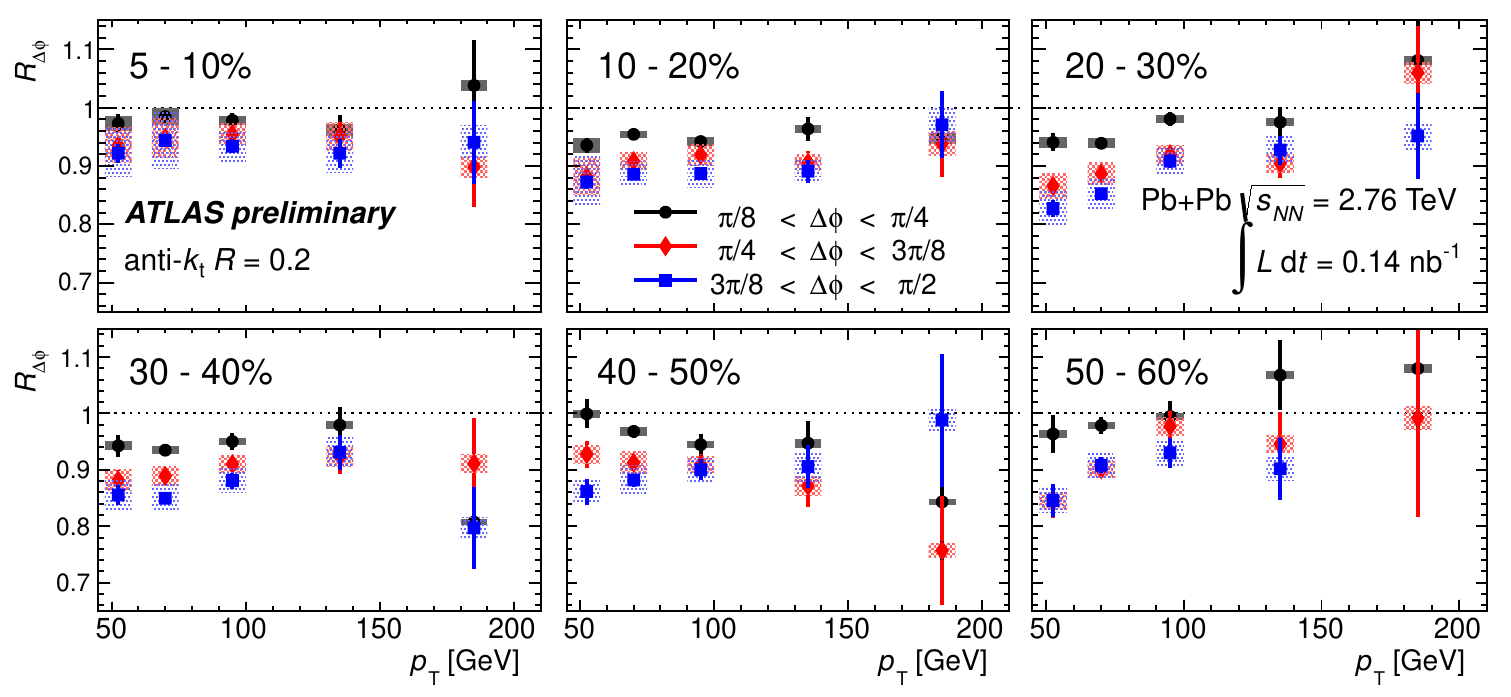}
}
\caption{
  Azimuthal dependence of the jet suppression measured in terms of 
$R_{\dphi}$ as a function of jet \pt\ for three bins of the angle
to the elliptic flow event plan, \dphi\ , evaluated relative to the $0 
< \dphi < \pi/8$. Error bars show statistical uncertainties, shaded 
boxes indicate systematic uncertainties \cite{azimuth}.
  }
\label{fig:rp}
\end{figure}

To get more insight into the dependence of the jet suppression on 
the path length the original parton would have traversed trough the 
medium the variation of inclusive jet yields as a function of 
azimuthal angle, \dphi\ , with respect to the elliptic flow event 
plane was measured. This variation can be quantified using the 
ratio of jet yield in one \dphi\ bin $i$ to another $j$, 
$R_{\dphi}$, defined by


\begin{equation}
R_{\dphi} = \frac{Y(\pt, \phi)|_{\dphi = \dphi_i} }
                  {Y(\pt, \phi)|_{\dphi = \dphi_j} } = 
\frac{ 
\frac{\fd^2 N(\pt, \phi)}{\fd \pt \fd \phi}\big|_{\dphi = \dphi_i }
}{
\frac{\fd^2 N(\pt, \phi)}{\fd \pt \fd \phi}\big|_{\dphi = \dphi_j } 
}
\end{equation}
  where the jet yield $Y(\pt, \phi)$
is measured in four \dphi\ bins spanning 
equidistantly the range $0 < \dphi < \pi/2$. The resulting $R_{\dphi}$ 
distributions plotted as a function of jet \pt\ are shown in 
Figure~\ref{fig:rp}. A clear reduction of the jet yield with increasing 
\dphi\ is observed. For example, in the 20-30\% centrality bin, jets 
produced in the direction perpendicular to the elliptic flow event plane ($3\pi/8 
< \dphi < \pi/2$) are suppressed by $\approx 15\%$ with respect to 
jets produced in the direction of the elliptic flow plane ($0 < \dphi < 
\pi/8$). Another way to quantify the azimuthal dependence of the jet 
suppression is to measure the elliptic flow parameter of jets, 
$v_2^{jet}$. The $v_2^{jet}$ was found to decrease with 
increasing centrality and to have a characteristic centrality 
dependence known from elliptic flow measurements of single particles 
\cite{flow}. The $v_2^{jet}$ reaches its maximum of $\approx 0.04-0.05$ 
in mid-central collisions ($N_{\mathrm{part}}=100-200$) for jets with 
$\pt = 45-60$~\GeV. More details on the azimuthal dependence of inclusive 
jet suppression can be found in Ref.~\cite{azimuth}.

\section{Photon-jet and $Z^{0}$-jet correlations}
\label{sec:gammajet}

The fact that the measured jet suppression at LHC is not an effect of 
initial state modifications was proven by the measurement of 
inclusive yields of prompt photons \cite{gamma} and inclusive yields 
of $W^{\pm}, Z^{0}$ vector bosons \cite{wboson, zboson} which do not exhibit any suppression 
in central heavy ion collisions. Since photons and vector bosons are 
insensitive to the colored medium they can provide means to calibrate 
the expected energy of quenched jet in a photon-jet system or in a 
$Z^{0}$-jet system. The quenching of the jet in the photon-jet sample 
can be quantified by measuring the jet to photon energy ratio, 
$x_{J\gamma} = \pt^{jet} / \pt^{\gamma}$. The left panel of 
Figure~\ref{fig:gaf} shows the integrated per-photon jet yield, 
$R_{J\gamma}$, evaluated as a function of the number of participants 
for data and MC. A clear suppression of the jet yield is seen in 
central compared to peripheral collisions. The measurement is 
performed for two different jet sizes, \RTwo\ and \RThree\ providing 
the same result within the systematic uncertainties. Qualitatively 
the same result has been obtained also from the $Z^{0}$-jet 
correlation measurements in which the $Z^{0}$ have been identified 
from the $e^+e^-$ and $\mu^+\mu^-$ decays. The same trend in the 
evolution of the jet suppression as a function of centrality has been 
observed despite to the limited statistics of less then forty 
$Z^{0}$-jet events. These results provide a starting point for 
precision measurements of the jet quenching which will be allowed by 
future high luminosity heavy ion runs at LHC. More details on 
$\gamma$-jet correlations are provided in Ref.~\cite{gammajet}, 
details on $Z^{0}$-jet correlations are provided in Ref.~\cite{zjet}.

\section{Heavy flavor suppression}
\label{sec:heavy}

\begin{figure}
\centerline{
\includegraphics[width=0.30\textwidth]{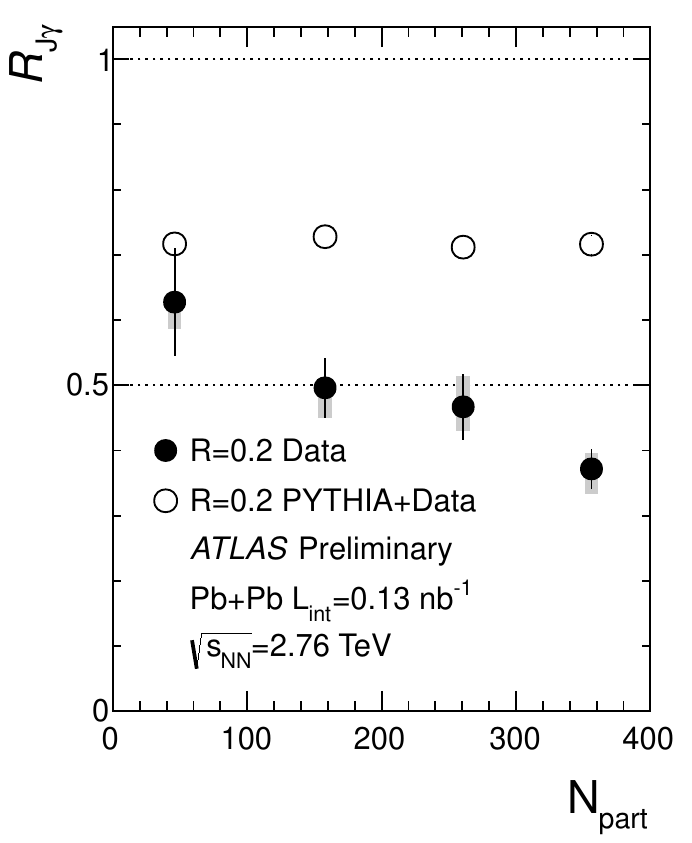}
\includegraphics[width=0.50\textwidth]{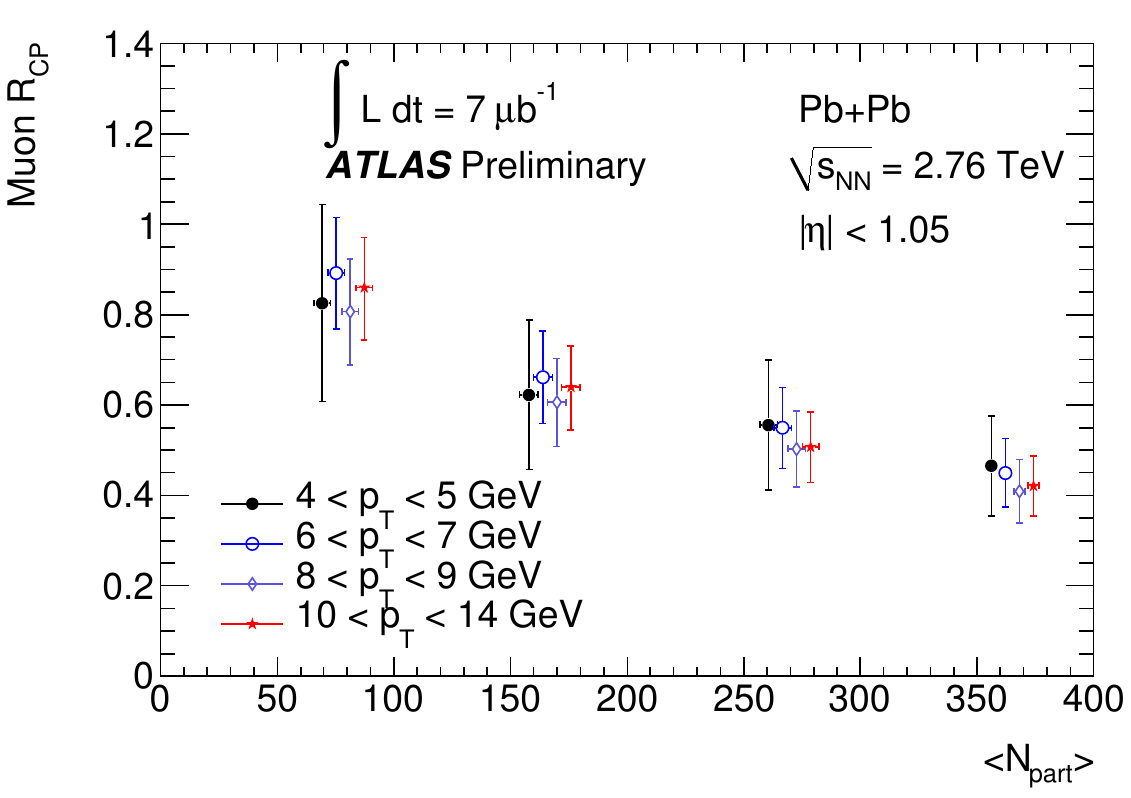}
}
\caption{
  {\it Left:}
  The integrated yield of jets per photon, $R_{J\gamma}$, calculated as a 
function of number of participating nucleons, \Npart . The kinematic cuts are: photon $60 < 
\pt^{\gamma} < 90$~\GeV, $|\eta^\gamma| < 1.3$, $\pt^{jet} > 25$~\GeV, 
$|\eta^{jet}| < 2.1$. The azimuthal distance between jet and photon is 
required to be $|\dphi_{J\gamma} | > 7\pi/8$. The error bars represent 
statistical errors, while the gray bands indicate the systematic 
uncertainties \cite{gammajet}, \cite{zjet}.
  {\it Right:}
  Muon \Rcp\ as a function of \Npart\ calculated for different bins in 
muon \pt . The sets of points for the different \pt\ bins are 
successively displaced by $\Delta\Npart = 6$ for clarity of 
presentation. The error bars represent combined statistical and 
systematic uncertainties \cite{heavyflavor}.
  }
\label{fig:gaf}
\end{figure}

To access the difference between the jet suppression of heavy 
quarks and light quarks the semi-leptonic decays of open heavy 
flavor hadrons into muons has been measured.
The measurement was performed over the muon transverse momentum 
range $4 < \pt < 14$~\GeV. Over this \pt\ range, muon production 
results predominantly from a combination of charm and bottom quark 
semi-leptonic decays. The differential yield of muons in a given 
centrality and \pt\ bin was compared to that in a peripheral bin 
through the central-to-peripheral ratio, \Rcp . The right panel of 
Figure~\ref{fig:gaf} shows a comparison of the muon \Rcp\ as a 
function of centrality for four different \pt\ intervals. One can 
see a smooth decrease in the \Rcp\ with increasing centrality. The 
suppression does not change with the muon \pt\ while the size of 
the suppression changes by nearly a factor of two between the most 
central and the most peripheral bins. A similar size of suppression 
between the yields of muons from semi-leptonic heavy quark decays 
and inclusive jet yields together with the \Rcp\ of muons being 
rather flat a as a function of \pt\ suggests no dramatic difference 
between the light and heavy quark response to the medium. More 
details on the heavy flavor suppression are provided in 
Ref.~\cite{heavyflavor}.

\section{Jet fragmentation}
\label{sec:frag}

One of the tools that allow precise comparison between the data and 
and theoretical models of jet quenching is the measurement of jet 
fragmentation.

\begin{figure}
\centerline{
\includegraphics[width=0.35\textwidth]{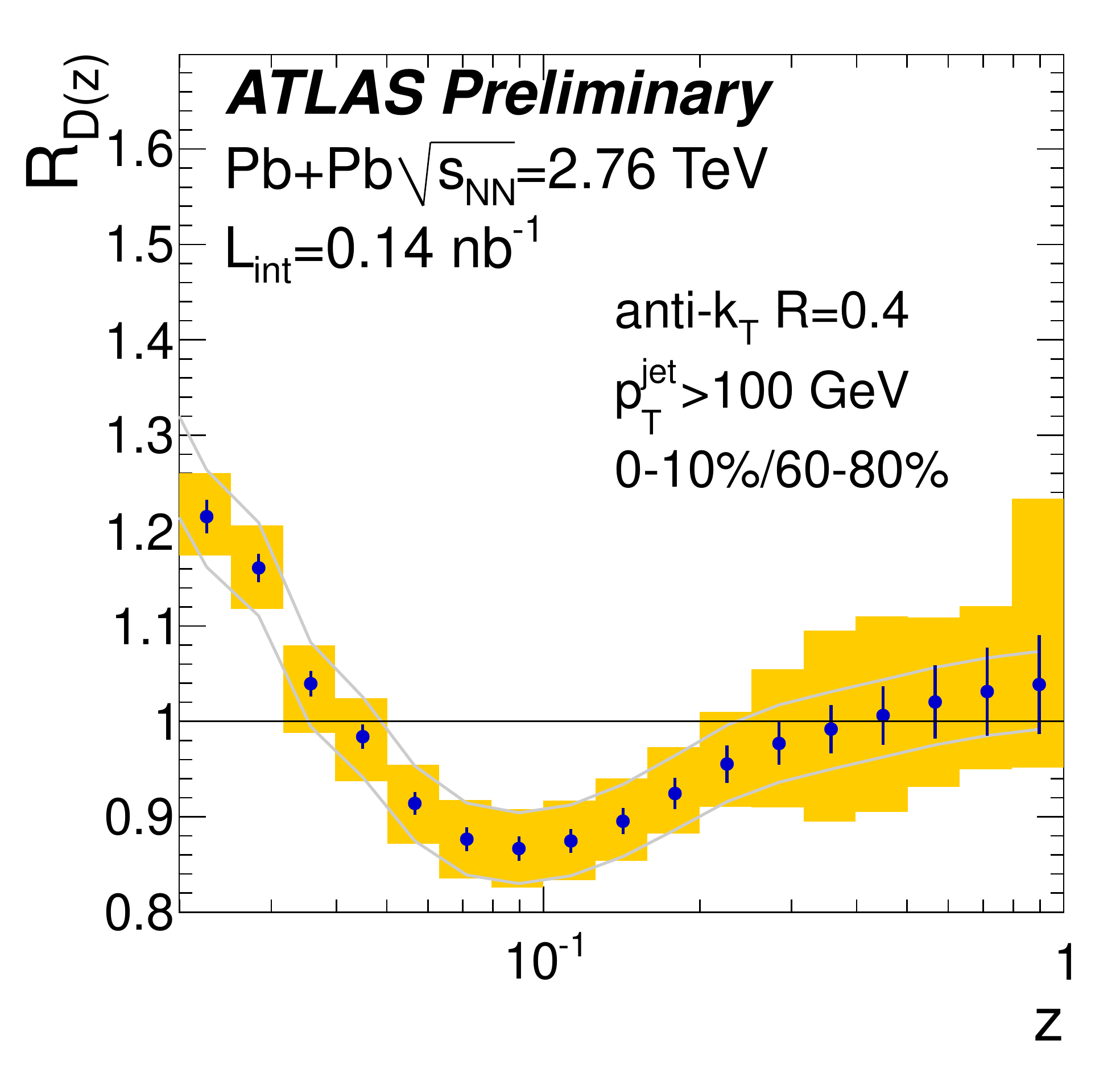}
\includegraphics[width=0.35\textwidth]{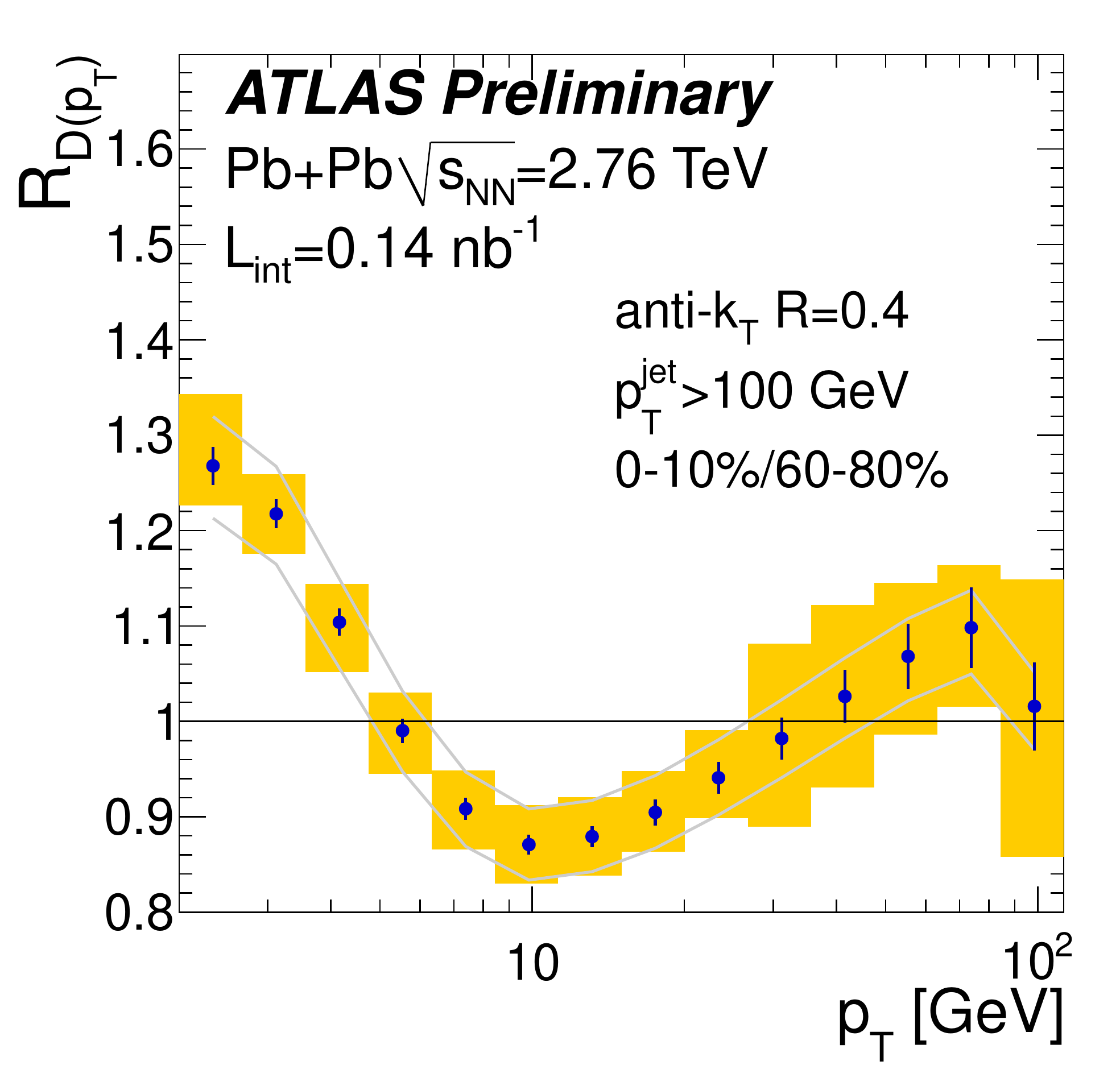}
}
\caption{ 
  {\it Left:} Ratio of fragmentation functions, $R_{D(z)}$. 
  {\it Right:} Ratio of \pt\ spectra of charged particles inside jets, 
$R_{D(\pt)}$. 
  Both $R_{D(z)}$ and $R_{D(\pt)}$ are ratios of 0-10\% central to 
60-80\% peripheral events evaluated for \RFour\ \antikt\ jets. The 
error bars on the data points show statistical uncertainties, the 
shaded bands indicate systematic uncertainties that are uncorrelated 
or partially correlated between points. The solid lines indicate 
systematic uncertainties that are 100\% correlated between points 
\cite{fragmentation}.
  }
\label{fig:frag}
\end{figure}

  The key question 
to address is how are the parton showers modified by the medium. 
To provide an experimental answer to this question we measure the \pt\ 
spectra of charged particles inside jets, $D(\pt)$, and the jet 
fragmentation function, $D(z)$, defined as

\begin{eqnarray}
D(\pt) &\equiv& \frac{1}{\Njet} \frac{1}{\varepsilon}
\frac{\Delta \Nch(\pt)}{\Delta \pt}\\
D(z) &\equiv& \frac{1}{\Njet} \frac{1}{\varepsilon}
\frac{\Delta \Nch(z)}{\Delta z}
\end{eqnarray}
  where \Njet\ represents the total number of measured jets in the 
given centrality bin, $\Delta \Nch (\pt)$ and $\Delta \Nch (z)$ 
represent the number of measured charged particles within $\Delta R = 
0.4$ of the jets in given bins of \pt\ and $z$, respectively, and 
$\varepsilon$ represents the MC-evaluated \pt\ and pseudorapidity dependent  reconstruction efficiency.
  The fragmentation variable, $z$, is defined as a ratio of charged 
particle momentum to the momentum of associated jet, 
  $z = \pt^{trk}/\pt^{jet} \cos \Delta R$, where $\Delta R$ is the distance 
  between the jet axis and a charged particle.
  The underlying event contribution to the $D(z)$ and $D(\pt)$ 
distributions is determined on the jet-by-jet basis and subtracted 
from the measured distributions. Jet fragmentation has been measured 
for three different jet radius parameters, $R=0.4, 0.3$, and 0.2 providing 
quantitatively the same result. The ratio of central to 60-80\% 
peripheral collisions of D(\pt) distributions, $R_{D(\pt)}$, and 
$D(z)$ distributions, $R_{D(z)}$, is calculated. The $R_{D(\pt)}$ and 
$R_{D(z)}$ has been measured for six centrality bins, 0-10\%, 
10-20\%, 20-30\%, 30-40\%, 40-50\%, and 50-60\%. The resulting 
distributions for \RFour\ jets and 0-10\% centrality bin are plotted 
in Figure~\ref{fig:frag}. The ratios show a reduction of fragment 
yield in all centrality bins relative to the 60-80\% bin at 
intermediate z values, $0.05 \lesssim z \lesssim 0.2$ and an 
enhancement in fragment yield for $z \lesssim 0.05$. No significant 
modification is seen in $D(z)$ or $D(\pt)$ distributions at high \pt\ 
or $z$. The reduction in the yield at intermediate $z$ decreases 
gradually from central to peripheral collisions. More details on the 
jet fragmentation measurement is provided in 
Ref.~\cite{fragmentation}.

The picture of the jet fragmentation at high-\pt\ is consistent with 
both the inclusive jet suppression and suppression of single charged 
particles reported in Ref.~\cite{trkspectra}. The charged particle \Rcp\ 
achieves a value of $\approx 0.6$, that is a similar value as the 
\Rcp\ of inclusive jets. Since the \Rcp\ of inclusive jets is rather 
flat as a function jet \pt\ and since all the measured charged 
particles at high-\pt\ should be also measured in jets we can expect 
no significant modification of yields of high-\pt\ fragments between 
central and peripheral collisions. This is indeed seen in the 
measurement of the jet fragmentation.

\section{Summary}
\label{sec:sum}

We have presented results of the measurement of inclusive jet 
suppression, azimuthal dependence of jet suppression, jet suppression 
in $\gamma$-jet and $Z^{0}$-jet system, suppression of heavy flavor, and 
jet fragmentation. We find that the inclusive jet yields are 
suppressed by approximately a factor of two in central compared to 
peripheral collisions. This suppression has only a weak dependence on 
the jet \pt . A significant dependence of the jet suppression on the 
jet size is seen -- with increasing jet radius the suppression is 
weaker. The jet suppression depends on the direction of the jet with 
respect to elliptic flow event plane. A maximal difference of $\approx 
15\%$ in the jet yields for jets produced in plane compared to jets 
produced perpendicular to the elliptic flow plane is observed. The jet 
suppression is also observed in $\gamma$-jet and $Z^{0}$-jet systems. 
The measurement of semi-leptonic decays of open heavy flavor hadrons 
to muons suggests a similar suppression of heavy quark jets and light 
quark jets. The jet fragmentation function and momentum spectra of 
charged particles produced inside jets are modified in central 
compared to peripheral collisions. The yield of low-$z$ fragments is 
enhanced whereas the yield of intermediate $z$ fragments is 
suppressed. No sizable difference of high-$z$ fragments is seen.




\bibliographystyle{elsarticle-num}
\bibliography{qm2012_new_v2}

\end{document}